\documentclass[prl,amssymb,amsmath,preprintnumbers,showpacs,twocolumn]{revtex4}

\usepackage{graphicx}
\usepackage{subfigure}
\usepackage{color}
\usepackage{verbatim}

\def\bs{\boldsymbol}

\begin{document}

\title{Generating shortcuts to adiabaticity in quantum and classical dynamics}

\author{Christopher Jarzynski}

\affiliation{Department of Chemistry and Biochemistry, and Institute for Physical Science and Technology,\\
University of Maryland, College Park, MD 20742 USA \\ \\
{\rm (Submitted for publication May 14, 2013)}}

\begin{abstract}
{\it Transitionless quantum driving} achieves adiabatic evolution in a hurry, using a counter-diabatic Hamiltonian to stifle non-adiabatic transitions.
Here this strategy is cast in terms of a generator of adiabatic transport, leading to a classical analogue: {\it dissipationless classical driving}.
For the single-particle piston, this approach yields simple and exact expressions for both the classical and quantal counter-diabatic terms.
These results are further generalized to even-power-law potentials in one degree of freedom.
\end{abstract}


\maketitle

According to the quantum adiabatic theorem~\cite{Messiah1966}, unitary evolution under a slowly time-dependent Hamiltonian $\hat H_0(t)$ closely tracks the instantaneous energy eigenstates $\vert n(t)\rangle$.
{\it Shortcuts to adiabaticity}~\cite{Chen2010b} are strategies for achieving the same result -- namely, evolving along the eigenstates of a time-dependent Hamiltonian -- without the requirement of slow driving.
This topic has received much attention in the past few years, see e.g.\ Refs.~\cite{Muga2009,Chen2010b,Chen2010c,Schaff2010,Stefanatos2010,Schaff2011,Bason2012,Ibanez2012,delCampo2012,delCampo2012b}, 
and has recently been reviewed by Torrontegui {\it et al}~\cite{Torrontegui2012}.

One such strategy, developed independently by Demirplak and Rice~\cite{Demirplak2003} and Berry~\cite{Berry2009}, employs a {\it counter-diabatic} Hamiltonian $\hat H_1(t)$, crafted to suppress transitions between energy eigenstates.
Consider a system that evolves under the Hamiltonian~\cite{Berry2009}
\begin{eqnarray}
\nonumber
\hat H(t) &=& \hat H_0(t) + i\hbar \, \sum_m \left(
\left\vert \partial_t m \right\rangle \left\langle m \right\vert - \left\langle m \vert \partial_t m\right\rangle \left\vert m \right\rangle \left\langle m \right\vert
\right) \\
\label{eq:berry}
&\equiv& \hat H_0(t) + \hat H_1(t) \, ,
\end{eqnarray}
where the sum is taken over the eigenstates $\vert m(t)\rangle$ of $\hat H_0(t)$, and $\vert \partial_t m\rangle \equiv \partial_t \vert m(t) \rangle$.
If such a system begins in the state $\vert n(0)\rangle$ at time $0$, then at all later times $t>0$ it will be found in the state $\vert n(t)\rangle$ (apart from an overall phase), even when the Hamiltonian is driven rapidly.
The term $\hat H_1(t)$ prevents the system from straying from the instantaneous eigenstate of $\hat H_0(t)$.

In this paper I argue that {\it transitionless quantum driving}~\cite{Berry2009} -- the strategy embodied by Eq.~\ref{eq:berry} -- is usefully framed in terms of a generator of adiabatic transport, $\hat{\bs\xi}$, which satisfies Eq.~\ref{eq:qconditions} below.
This perspective suggests a natural extension to classical systems, which might be called {\it dissipationless classical driving}.
Moreover, the framework developed in this Letter offers an alternative approach to constructing the counter-diabatic Hamiltonian $H_1(t)$.
When applied to the paradigmatic example of a particle in a one-dimensional box~\cite{delCampo2012b}, this approach yields simple expressions for the counter-diabatic term for both the classical and the quantal versions of this problem (Eqs.~\ref{eq:boxHt}, \ref{eq:boxHtq}).
These solutions are readily generalized to potentials of the form $V(q) \propto q^b$, where $b>0$ is an even integer (Eqs.~\ref{eq:chtb}, \ref{eq:qhtb}).
In a recent posting to {\it arXiv.org}, Deng {\it et al}~\cite{Deng2013} have independently developed the idea of dissipationless classical driving from a somewhat different perspective.

To begin, let $\hat H_0$ be an explicit function of external parameters $\bs\lambda = (\lambda_1,\cdots \lambda_N)$, with eigenstates $\vert n(\bs\lambda) \rangle$ and eigenvalues $E_n(\bs\lambda)$.
Given a schedule ${\bs\lambda}(t)$ for varying these parameters, Eq.~\ref{eq:berry} takes the form
\begin{equation}
\label{eq:implicit}
\hat H(t) = \hat H_0({\bs{\lambda}}(t)) + \dot {\bs\lambda} \cdot \hat{\bs\xi}({\bs\lambda}(t)) \, ,
\end{equation}
where $\hat{\bs\xi}({\bs\lambda}) = (\hat\xi_1, \cdots \hat\xi_N )$ is a vector of Hermitian operators:
\begin{equation}
\label{eq:berryxidef}
\hat{\bs\xi}({\bs\lambda}) = i \hbar \sum_m \left(
\left\vert {\bs\nabla} m \right\rangle \left\langle m \right\vert - \left\langle m \vert {\bs\nabla} m\right\rangle \left\vert m \right\rangle \left\langle m \right\vert
\right) 
\end{equation}
with $\vert {\bs\nabla} m\rangle \equiv \partial_{\bs\lambda} \vert m({\bs\lambda}) \rangle$ and $\dot{\bs\lambda} \equiv {\rm d}{\bs\lambda}/{\rm d}t$.

Let us now view $\hat{\bs\xi}(\bs\lambda)$ as a {\it generator} that associates infinitesimal displacements in parameter space, ${\bs\lambda} \rightarrow {\bs\lambda} + \delta {\bs\lambda}$, with displacements in Hilbert space, $\vert\psi\rangle \rightarrow \vert\psi\rangle + {\vert\delta\psi\rangle}$, according to the rule
\begin{equation}
\label{eq:qrule}
i\hbar \, \vert\delta\psi\rangle = \delta\bs\lambda \cdot \hat{\bs\xi} \, \vert\psi\rangle \, .
\end{equation}
When applied to an eigenstate of $\hat H_0({\bs\lambda})$, this prescription generates the displacement
\begin{equation}
\label{eq:qadiabaticDisplacement}
\vert n({\bs\lambda})\rangle \rightarrow
\left( 1 + \frac{1}{i\hbar} \delta{\bs\lambda}\cdot\hat{\bs\xi} \right) \vert n({\bs\lambda})\rangle =
e^{i\,\delta{\bs\lambda}\cdot {\bs A}_n} \vert n({\bs\lambda} + \delta{\bs\lambda})\rangle \, ,
\end{equation}
(to first order in $\delta{\bs\lambda}$),
as follows from Eq.~\ref{eq:berryxidef}, with ${\bs A}_n({\bs\lambda}) = i \langle n \vert {\bs\nabla} n\rangle$.
If we start in a state $\vert n({\bs\lambda}_0)\rangle$ and apply Eq.~\ref{eq:qrule} stepwise along a curve ${\bs\lambda}_s$ in parameter space, then the wavefunction gets transported along the curve $e^{i\varphi_s} \vert n({\bs\lambda}_s)\rangle$, with the phase given by the line integral of ${\bs A}_n({\bs\lambda})$.
Thus $\hat{\bs\xi}$ generates a unitary flow in Hilbert space, induced by the variation of the parameters, which escorts the system along eigenstates of $\hat H_0({\bs\lambda})$.

The flow described above is parametric rather than temporal.
Now consider evolution under the time-dependent Schr\" odinger equation, with $\hat H(t)$ given by Eq.~\ref{eq:implicit}.
During an infinitesimal time interval $\delta t$, a wave function $\vert \psi\rangle$ evolves to:
\begin{equation}
\left( 1 + \frac{1}{i\hbar} \hat H \, \delta t \right) \vert\psi\rangle =
\vert\psi\rangle + \frac{1}{i\hbar} \delta t \, \hat H_0 \vert\psi\rangle + \frac{1}{i\hbar} \, \delta{\bs\lambda}\cdot\hat{\bs\xi} \vert\psi\rangle \, .
\end{equation}
If we set $\vert\psi\rangle = \vert n({\bs\lambda})\rangle$, the effects of the terms $\hat H_0$ and $\dot{\bs\lambda}\cdot\hat{\bs\xi}$ are simple to state:
the first produces the familiar dynamical phase associated with quantal time evolution, 
while the second directly couples changes in ${\bs\lambda}$ to displacements in Hilbert space, in a way that enforces the adiabatic constraint (Eq.~\ref{eq:qadiabaticDisplacement}).
Transitionless quantum driving is achieved with $\hat H(t)$, precisely because $\hat{\bs\xi}$ has been fashioned to guide systems along eigenstates of $\hat H_0$ under parametric changes.

The generator $\hat{\bs\xi}$ defined by Eq.~\ref{eq:berryxidef} can alternatively be specified by the conditions,
\begin{subequations}
\label{eq:qconditions}
\begin{eqnarray}
\label{eq:qc1}
\bigl[ \hat{\bs\xi} , \hat H_0 \bigr] &=& i\hbar \left ( {\bs\nabla} \hat H_0 - {\rm diag} ( {\bs\nabla} \hat H_0 ) \right) \\
\label{eq:qc2}
\bigl\langle n \bigl\vert \hat{\bs\xi} \bigr\vert n \rangle &=& 0 \quad ,
\end{eqnarray}
\end{subequations}
where ${\rm diag} ( {\bs\nabla} \hat H_0 ) = \sum_m \vert m\rangle \langle m\vert{\bs\nabla}\hat H_0\vert m\rangle \langle m\vert$.
Eq.~\ref{eq:qc1} determines the off-diagonal elements of $\hat{\bs\xi}$, as can be seen by applying the operation $\langle m \vert \cdots \vert n \rangle$ to both sides; and Eq.~\ref{eq:qc2} sets the diagonal elements.
The identity $\langle m \vert {\bs\nabla} n \rangle = \langle m \vert {\bs\nabla} \hat H_0 \vert n \rangle / (E_n-E_m)$~\cite{Berry2009} establishes the equivalence of the two definitions of $\hat{\bs\xi}$ (Eqs.~\ref{eq:berryxidef} and \ref{eq:qconditions}).

Eq.~\ref{eq:qconditions} suggests an avenue for developing a classical counterpart of transitionless quantum driving.
Consider a classical Hamiltonian in one degree of freedom, $H_0(z;{\bs\lambda})$, where $z=(q,p)$ specifies a point in two-dimensional phase space.
Assume further that the {\it energy shells} (level surfaces of $H_0$) form closed, simple loops in phase space,  identified by their energies $E = H_0(z;{\bs\lambda})$.
If
\begin{equation}
\label{eq:Omegadef}
\Omega(E,{\bs\lambda}) \equiv  \int {\rm d} z \, \theta \left[ E - H_0(z;{\bs\lambda}) \right]
\end{equation}
denotes the volume of phase space enclosed by the energy shell $E$, then the observable
\begin{equation}
\label{eq:omegadef}
\omega(z;{\bs\lambda}) \equiv \Omega\bigl( H_0(z;{\bs\lambda}) , {\bs\lambda} \bigr)
\end{equation}
is an adiabatic invariant~\cite{Goldstein1980}: when the system evolves under Hamilton's equations as the parameters are varied infinitely slowly, the value of $\omega(z(t);{\bs\lambda}(t))$ remains constant along the trajectory $z(t)$.
For later convenience, let angular brackets denote a microcanonical average:
\begin{equation}
\label{eq:microcanonicalAverage}
\left\langle\cdots\right\rangle_{E,{\bs\lambda}} \equiv 
\frac{1}{\partial_E\Omega} \int {\rm d}z \, \delta (E - H_0) \cdots
\end{equation}
Inverting $\Omega(E,{\bs\lambda})$ to define $E(\Omega,{\bs\lambda})$, we obtain
\begin{equation}
\label{eq:cyclic}
{\bs\nabla} E(\Omega,{\bs\lambda}) = - \frac{ {\bs\nabla} \Omega(E,{\bs\lambda}) } { \partial_E\Omega(E,{\bs\lambda}) }
= \left\langle {\bs\nabla} H_0 \right\rangle_{E,{\bs\lambda}}
\end{equation}
using Eqs.~\ref{eq:Omegadef} and \ref{eq:microcanonicalAverage}, and the cyclic identity of partial derivatives.
With these elements in place, let us construct a Hamiltonian under which the value of $\omega$ is preserved {\it exactly}, again using a counter-diabatic term $\dot{\bs\lambda}\cdot{\bs\xi}$ to enforce adiabatic discipline (Eq.~\ref{eq:cHoft}).

A semiclassical counterpart of Eq.~\ref{eq:qconditions} is given by~\cite{Jarzynski1995b}
\begin{subequations}
\label{eq:cconditions}
\begin{eqnarray}
\label{eq:cc1}
\bigl\{ {\bs\xi} , H_0 \bigr\} &=&  {\bs\nabla} H_0 - \left\langle {\bs\nabla} H_0 \right\rangle_{H_0,{\bs\lambda}} \equiv {\bs\nabla} \tilde H_0  \\
\label{eq:cc2}
\bigl\langle {\bs\xi}  \bigr\rangle_{E,{\bs\lambda}} &=& 0 \quad ,
\end{eqnarray}
\end{subequations}
where $\{\cdot,\cdot\}$ denotes the Poisson bracket~\footnote{
$\{A,B\} = (\partial A/\partial q)\,(\partial B/\partial p) - (\partial A/\partial p)\,(\partial B/\partial q)$
}.
Using Eq.~\ref{eq:cyclic}, Eq.~\ref{eq:cc1} can be rewritten in the simpler form
\begin{equation}
\label{eq:simplecc1}
\{ {\bs\xi} , \omega \} = {\bs\nabla} \omega \, .
\end{equation}
By analogy with the quantal case, let us treat ${\bs\xi}(z;{\bs\lambda})$ as a generator that converts displacements in parameter space, ${\bs\lambda} \rightarrow {\bs\lambda} + \delta {\bs\lambda}$, to displacements in phase space, $z \rightarrow z + \delta z$, according to the rule
\begin{equation}
\label{eq:crule}
\delta z = \delta{\bs\lambda} \cdot \{ z, {\bs\xi} \} \, .
\end{equation}
Under this prescription, ${\bs\xi}(z;{\bs\lambda})$ generates a canonical flow in phase space, induced by the variation of ${\bs\lambda}$, that preserves the value of $\omega$:
\begin{equation}
\label{eq:cProof}
\begin{split}
\omega \bigl( z &+ \delta z ; {\bs\lambda} + \delta{\bs\lambda} \bigr) - \omega \bigl( z ; {\bs\lambda} \bigr) \\
&= \frac{\partial\omega}{\partial z} \delta z + {\bs\nabla} \omega \cdot \delta{\bs\lambda} \\
&= \bigl( \{ \omega,{\bs\xi} \} + {\bs\nabla} \omega \bigr) \cdot \delta{\bs\lambda} = 0 \, .
\end{split}
\end{equation}
using Eqs.~\ref{eq:simplecc1} and \ref{eq:crule}.
Thus the transformation $z \rightarrow z + \delta{\bs\lambda} \cdot \{ z, {\bs\xi} \}$ maps points from a single energy shell of $H_0(z;{\bs\lambda})$, onto the energy shell of $H_0(z;{\bs\lambda}+\delta{\bs\lambda})$ that encloses the same phase space volume.

Now consider a trajectory $z(t)$ evolving under Hamilton's equations, $\dot z = \{z,H\}$, with
\begin{equation}
\label{eq:cHoft}
H(z,t) = H_0(z;{\bs\lambda}(t)) + \dot{\bs\lambda} \cdot {\bs\xi}(z,{\bs\lambda}(t)) \, .
\end{equation}
Again using Eq.~\ref{eq:simplecc1}, we obtain
$({\rm d}/{\rm d}t) \, \omega(z(t);{\bs\lambda}(t)) = 0$.
As advertised, the counter-diabatic term $\dot{\bs\lambda} \cdot {\bs\xi}$
ensures that the adiabatic invariant is conserved exactly.

It is useful to consider this process in terms of an ensemble of trajectories.
Imagine a collection of initial conditions sampled from an energy shell $E_0$ of $H_0(z;{\bs\lambda}(0))$.
At any later time $t>0$, the trajectories that evolve from these initial conditions, under the Hamiltonian $H(z,t)$, will populate a single energy shell $E(t)$ of $H_0(z;{\bs\lambda}(t))$, specifically the {\it adiabatic energy shell} enclosing the same volume of phase space as the initial shell.
If we picture the adiabatic shell as a closed loop that deforms as the parameters ${\bs\lambda}$ are varied with time, then in Eq.~\ref{eq:cHoft} $H_0$ generates motion around this loop, and $\dot{\bs\lambda}\cdot{\bs\xi}$ adjusts each trajectory so that it remains on the shell.

\begin{figure}[tbp]
\includegraphics[trim = 1in 2in 0in 0in , scale=0.4,angle=0]{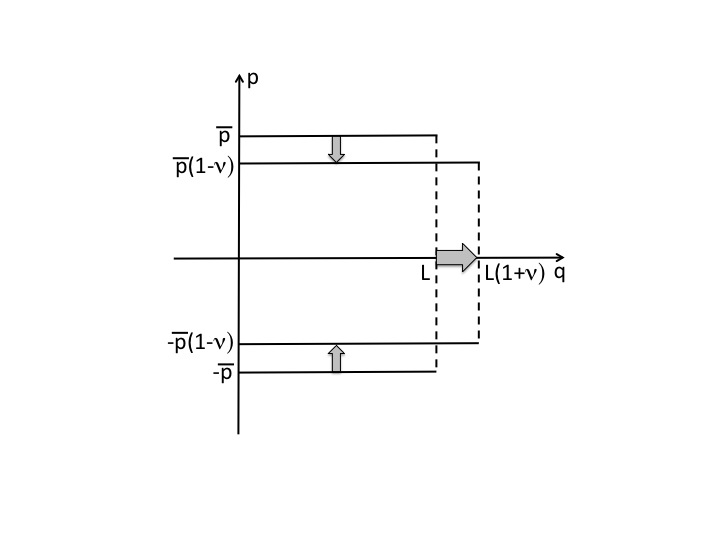}
\caption{
Energy shells for a particle in a one-dimensional box (Eq.~\ref{eq:cboxHam}).
One shell is shown as a pair of parallel line segments of length $L$, at momenta $\pm \overline p$.
When the box length is increased infinitesimally, the new adiabatic energy shell is obtained by stretching in the $q$ direction and contracting in $p$.
}
\label{fig:1dbox_energyShells}
\end{figure}

The fact that the transformation $z\rightarrow z+\delta{\bs\lambda} \cdot \{ z, {\bs\xi} \}$ maps an energy shell of $H_0(z;{\bs\lambda})$ onto an energy shell of $H_0(z;{\bs\lambda} + \delta{\bs\lambda})$ provides intuition that  can be exploited in constructing ${\bs\xi}(z;{\bs\lambda})$.
Consider a particle of mass $m$ inside a one-dimensional box with hard walls at $q=0$ and $q=L$, as described by a Hamiltonian
\begin{equation}
\label{eq:cboxHam}
H_0(z;L) = \frac{p^2}{2m} + V_{\rm box}(q;L) \, ,
\end{equation}
where $V_{\rm box}(q;L)$ is zero inside the box, and ``infinite'' outside~\footnote{$V_{\rm box}(q;L)$ can be considered as the limiting case of a potential well with soft walls.}.
We wish to construct a counter-diabatic term for processes in which the box length $L$ is varied with time.
To this end, note that an energy shell $E$ consists of two line segments in phase space,
forming the upper and lower edges of a rectangle of length $L$, width $2\overline p = \sqrt{8mE}$, and volume $\Omega = 2\overline p L$ (Fig.~\ref{fig:1dbox_energyShells}).
For a slightly larger box size, the energy shell enclosing the same volume of phase space is described by a rectangle of length $L(1+\nu)$ and width $2\overline p (1- \nu)$, where $\nu = \delta L/L$ is the fractional increase in the box size.
These two energy shells are related by linear scaling;
the first is mapped onto the second by a linear expansion along $q$ and a compensating contraction along $p$:
\begin{equation}
\label{eq:deltas}
q \rightarrow q \, (1+\nu) \quad,\quad p \rightarrow p \, (1-\nu)
\, .
\end{equation}
We can work backward from this canonical transformation to solve for its generator, by setting
$(\nu q, -\nu p) \equiv \delta z = \delta L \{ z,\xi \}$ (Eq.~\ref{eq:crule}).
This produces the pair of equations $q/L = \partial\xi/\partial p$ and $p/L = \partial\xi/\partial q$,
whose solution is $\xi = qp/L$, with the constant of integration set by Eq.~\ref{eq:cc2}.
With Eq.~\ref{eq:cHoft} this finally leads to
\begin{equation}
\label{eq:boxHt}
H(z,t) = H_0(z;L) + \frac{\dot L}{L} qp \quad,\quad L= L(t)
\, .
\end{equation}
Under this time-dependent Hamiltonian, the adiabatic invariant $\omega(q,p;L) = 2\vert p \vert L$ is conserved exactly, for any choice of the schedule $L(t)$.
This can be verified by inspection of Hamilton's equations, with care devoted to the collisions between the particle and the moving wall.

To gain intuition for the counter-diabatic term in Eq.~\ref{eq:boxHt}, note that by Hamilton's equations,
\begin{equation}
\label{eq:qdotbox}
\dot q = \frac{\partial H}{\partial p} = \frac{p}{m} + \frac{\dot L}{L} q
\, .
\end{equation}
The last term produces a linear scaling whose effect is most easily pictured by imagining a lattice of particles at rest ($p=0$), distributed  at equally spaced intervals within the box.
Under Eq.~\ref{eq:qdotbox} this lattice is uniformly stretched or contracted along with the box length.
Now, more generally, consider the evolution of a gas of independent particles initially distributed uniformly within the box, with an arbitrary distribution of momenta.
Even as the box length $L$ is varied arbitrarily with time, the gas remains distributed uniformly throughout the box:
the counter-diabatic term in Eq.~\ref{eq:boxHt} perfectly suppresses shock waves by uniformly expanding or compressing the gas.

Let us now use the classical solution $\xi=qp/L$ as a starting point for seeking the corresponding quantal generator.
Since the operators $\hat q$ and $\hat p$ do not commute, a natural first guess is
\begin{equation}
\label{eq:firstGuess}
\hat\xi(L) = \frac{\hat q \hat p + \hat p \hat q}{2L} \, .
\end{equation}
As luck would have it, an explicit evaluation confirms that this choice satisfies Eq.~\ref{eq:qadiabaticDisplacement}:
\begin{equation}
\label{eq:lucky}
\left( 1 + \frac{1}{i\hbar} \delta L \, \hat\xi \right)
\sqrt{\frac{2}{L}} \sin \left( \frac{n\pi q}{L} \right) = \sqrt{\frac{2}{L+\delta L}} \sin \left( \frac{n\pi q}{L+\delta L} \right)
\end{equation}
with $A_n(L) = i \langle n \vert \partial_L n\rangle = 0$.
We can then immediately write down a Hamiltonian
\begin{equation}
\label{eq:boxHtq}
\hat H(t)
= -\frac{\hbar^2}{2m} \frac{\partial ^2}{\partial q^2} + V_{\rm box}(q;L) + \frac{\dot L}{2L} \frac{\hbar}{i}
\left( q \frac{\partial}{\partial q} + \frac{\partial}{\partial q} q \right) \, ,
\end{equation}
for which the wave function
\begin{equation}
\label{eq:boxPsiexact}
\psi(q,t) = \sqrt{\frac{2}{L}} \sin\left(\frac{n\pi q}{L}\right) \exp \left( -\frac{i}{\hbar} \int_0^t {\rm d}t^\prime \, \frac{n^2\pi^2\hbar^2}{2mL^2} \right)
\end{equation}
is an exact solution of the Schr\" odinger equation (as verified by inspection) for arbitrary $L(t)$.
Eq.~\ref{eq:boxHtq} contains precisely the counter-diabatic term needed to achieve transitionless quantum driving for this example.

Eq.~\ref{eq:lucky} can be written more generally as
\begin{equation}
\label{eq:stretch}
\left( 1 + \frac{1}{i\hbar} \delta L \, \hat\xi \right) \psi(q) =
\sqrt{ \frac{1}{s} } \,\, \psi\left( \frac{q}{s} \right)
\quad,\quad
s = \frac{L + \delta L}{L}
\end{equation}
where $\psi(q)$ is any differentiable wavefunction.
In other words, the operator $\exp(\delta L\,\hat\xi/i\hbar)$ stretches $\psi(q)$ linearly, while preserving its norm, $\int \vert\psi\vert^2 {\rm d}q$.
In the momentum representation, this operator contracts the wavefunction: increasing the local wavelength reduces the local momentum.
Thus the action of $\hat\xi(L)$ in Hilbert space mimics that of $\xi(z;L)$ in phase space (Eq.~\ref{eq:deltas}).

It should be clear that, in this particular example, transitionless quantum driving (suppression of non-adiabatic transitions) and dissipationless classical driving (suppression of shock waves) are achieved by virtue of the scaling relation that holds among the adiabatic eigenstates or energy shells.
E.g.\ to follow the eigenstate $\phi_n(q;L) \propto \sin(n\pi q/L)$, the wavefunction needs simply to be stretched or contracted appropriately (Eq.~\ref{eq:stretch}).
Similar scaling relations~\cite{Landau_mechanicalSimilarity} apply to all even-power-law potentials in one degree of freedom, represented by the classical Hamiltonian
\begin{equation}
\label{eq:h0b}
H_0(z;L) = \frac{p^2}{2m} + \epsilon \left( \frac{q}{L} \right)^b
\end{equation}
or its quantal counterpart.
Here $\epsilon > 0$ sets the energy scale and $b \in \{2,4,6,\cdots\}$.
The adiabatic invariant is
\begin{equation}
\omega(z;L) = c L H_0^{\frac{1}{2}+\frac{1}{b}}
= c L H_0^{\frac{1}{2\mu}}
\quad,\quad
\mu = \frac{b}{b+2}
\, ,
\end{equation}
where $c$ is a constant~\footnote{Explicitly, $c =\sqrt{8\pi m} \, \epsilon^{-1/b} \,  \Gamma(1 + \frac{1}{b}) / \Gamma(\frac{3}{2} + \frac{1}{b})$.}.
The change $L \rightarrow L + \delta L$ induces a change in the adiabatic energy shell that 
is described by the linear canonical transformation given by Eq.~\ref{eq:deltas}, only now with
$\nu = \mu \, \delta L/L$.
Proceeding as before, we arrive at the Hamiltonian
\begin{equation}
\label{eq:chtb}
H(z,t) = H_0(z;L) + \frac{b}{b+2} \frac{\dot L}{L} qp \, ,
\end{equation}
under which the value of $\omega = cLH_0^{1/2\mu}$ is conserved exactly, for any schedule $L(t)$.
For the quantal version of Eq.~\ref{eq:h0b}, the eigenstates satisfy
\begin{equation}
\phi_n(q;L) = \sqrt{\frac{1}{L^\mu}} \, \phi_n\left( \frac{q}{L^\mu} ; 1 \right)
\, ,
\end{equation}
and transitionless quantum driving is achieved with the Hamiltonian
\begin{equation}
\label{eq:qhtb}
\hat H(t) = \hat H_0(L) + \frac{b}{b+2}  \frac{\dot L}{2L} \frac{\hbar}{i}
\left( q \frac{\partial}{\partial q} + \frac{\partial}{\partial q} q \right)
\, .
\end{equation}
This result is valid for all $b \in \{2,4,6,\cdots\}$, even when explicit expressions for the energy eigenstates are unavailable.
For $b=2$ and $b\rightarrow\infty$ this problem reduces to a harmonic oscillator and a particle in a box (with walls at $\pm L$), respectively.
Indeed, for the harmonic oscillator Eq.~\ref{eq:qhtb} was obtained earlier by Muga {\it et al}~\cite{Muga2010}, using ladder operators $\hat a$ and $\hat a^\dagger$ to evaluate Eq.~\ref{eq:berry}.

For other potentials in one degree of freedom, the generator ${\bs\xi}$ satisfies
\begin{equation}
\label{eq:trajectorySolution}
{\bs\xi}(z_b;{\bs\lambda}) - {\bs\xi}(z_a;{\bs\lambda}) = \int_a^b {\rm d}t \, {\nabla}\tilde H_0(z(t);{\bs\lambda})
\, ,
\end{equation}
where $z_a$ and $z_b$ are two points on the same energy shell of $H_0(z;{\bs\lambda})$, and $z(t)$ is a trajectory that evolves under $H_0$ from $z_a$  to $z_b$.
[Eq.~\ref{eq:trajectorySolution} follows by combining Eq.~\ref{eq:cc1} with the Hamiltonian identity $({\rm d}/{\rm d}t) {\bs\xi}(z(t);{\bs\lambda}) = \{ {\bs\xi},H_0 \}$.]
Thus by integrating ${\bs\nabla}\tilde H_0$ along a trajectory for one period of motion, the function ${\bs\xi}$ can be determined for all points on the energy shell, up to an additive constant that in turn is set by Eq.~\ref{eq:cc2}.

The situation becomes more complicated if we drop the assumption that the energy shells of $H_0$ are simple, closed loops.
For instance, if $V(q;{\bs\lambda})$ describes a double-well potential, then the variation of ${\bs\lambda}$ may cause an energy shell to change its topology from a single loop to a double loop (or vice-versa), upon passing through a separatrix.
The adiabatic invariance of $\omega$ then breaks down~\cite{Cary1986,Tennyson1986,Hannay1986}.
It remains an open question how the framework developed in this paper might extend to such situations.

For classical systems with $N\ge 2$ degrees of freedom, if $H_0$ is integrable then it may be possible to repeat the analysis of Eqs.~\ref{eq:Omegadef} - \ref{eq:cHoft} in terms of action-angle variables, with a generator ${\bs\xi}_i$ associated with each action-angle pair.
At the other extreme, if $H_0$ is ergodic then the existence of a solution of Eq.~\ref{eq:cc1} implies that the energy shells of $H_0(\vec z;{\bs\lambda})$ can be mapped by a canonical transformation to those of $H_0(\vec z;{\bs\lambda}+\delta{\bs\lambda})$~\cite{Jarzynski1995b}.
This condition is not generically satisfied when $N\ge 2$, but if it is, then ${\bs\xi}$ is simply the generator of this transformation, and dissipationless driving is achieved with $H_0 + \dot{\bs\lambda}\cdot{\bs\xi}$.

Finally, while the approach described in this paper applies to quantum and classical dynamics, analogous ingredients arise 
in a proposed statistical-mechanical method for the efficient estimation of free energy differences, where a counter-diabatic {\it metric scaling}~\cite{Miller2000} or {\it flow field}~\cite{Vaikuntanathan2008} is constructed to reduce or eliminate irreversibility in numerical simulations of finite-time processes.

It is a pleasure to acknowledge stimulating discussions with M.V.\ Berry, S.\ Deffner, D.\ Mandal and V.\ Yakovenko, and support from the National Science Foundation (USA) under grant DMR-1206971.

\bibliography{CJ_references}

\end{document}